\documentclass[aps,prb,twocolumn]{revtex4-1}
\usepackage{graphicx}

\begin{document}

\title{Ni-doped Epitaxial Graphene Monolayer on the Ni(111) Surface}

\author{S. L. Kovalenko}
\author{T. V. Pavlova}
\author{B. V. Andryushechkin}
\author{G. M. Zhidomirov}
\author{K. N. Eltsov}
\affiliation{Prokhorov General Physics Institute of the Russian Academy of Sciences, Moscow, Russia}

\begin{abstract}

Nickel-doped graphene has been synthesized from propylene on a Ni(111) surface and studied using scanning tunneling microscopy (STM) and density functional theory (DFT). It is established that nickel centers are formed during graphene synthesis on the Ni(111) surface by both chemical vapor deposition (CVD) and temperature-programmed growth (TPG); apparently, they are always present in graphene synthesized on Ni(111). The centers are observed in STM images as single defects or defect chains and identified by DFT calculations as Ni atoms in carbon bivacancies. These nickel atoms are positively charged and may be of interest for single-atom catalysis. The incorporated Ni atoms should remain in graphene after the detachment from the substrate since they bound more strongly with carbon atoms in graphene than with substrate nickel atoms.

\end{abstract}

\maketitle

\section{\label{sec:intro}Introduction}

Defects in graphene are known to affect significantly its properties. In particular, their presence may cause the shift of Fermi level, opening locally the energy gap. Moreover, defects can influence the carrier mobility, conductivity, magnetic and mechanical properties of graphene; and play the role of active centers in chemical reactions \cite{2011Banhart}. Doping graphene with nitrogen and boron also affects the Fermi level position (these processes can be considered as p-doping \cite{2013Cattelan, 2013Gebhardt} and n-doping \cite{2011Usachov, 2012Velez-Fort}, respectively), leads to the energy gap opening ($\approx$0.2 eV for nitrogen \cite{2011Usachov}), and changes the carrier mobility in graphene \cite{2010Guo, 2013Wang}.

Graphene doped with individual metal atoms has recently been studied for use as a single atom catalyst (SAC), in particular, in the hydrogen evolution reaction (HER) \cite{2015Qiu,2018Zhang} and oxygen reduction reactions (ORR) \cite{2018Zhang,2019Mao}. Moreover, for single-atom catalysis, not only free graphene is interesting, but also graphene on a substrate (i.e. Ni(111) \cite{2019Mao}), since the presence of a substrate allows one to change the activation barrier of reactions.

The graphene/Ni(111) system attracts a lot of interest (see review \cite{2014Dahal}) due to the small (1.2\%) lattices mismatch that potentially makes possible the fabrication  of the large area graphene crystals of high quality. Many researchers \cite{2012Jacobson,2014Bianchini, 2015Smerieri, 2013Patera, 2019Carnevali, 2015Garcia-Lekue,2017Parreiras,2017Kovalenko} observed characteristic objects in scanning tunneling microscopy (STM) images of graphene synthesized on Ni(111), which looked much brighter than graphene structural defects.

The observation of such features (both individual objects and chains) was reported for graphene synthesized on Ni(111) by chemical vapor deposition (CVD) from toluene  \cite{2012Jacobson} or ethylene \cite{2014Bianchini}, although, their structure was not considered. Similar objects were observed in graphene synthesized by CVD from ethylene and interpreted as possible impurities (molecules from the residual atmosphere of ultrahigh vacuum chamber) deposited on graphene structural defects \cite{2015Smerieri}. In Ref. \cite{2013Patera}, bright objects with a concentration of 1--2$\%$ were observed in graphene formed on the Ni(111) surface from ethylene by CVD at the temperature range of 400--500$^{\circ}$C. The interpretation assumed a possible replacement of carbon atoms by nickel ones, however, the structure details were not discussed. Recently, it has been shown that the incorporation of nickel atoms into the graphene lattice does occur during CVD synthesis of graphene on Ni(111), moreover, these atoms are visualized in STM as bright points \cite{2019Carnevali}.

Note that similar bright objects were also observed on graphene islands synthesized by temperature programmed growth (TPG) on the Ni(111) surface from ethylene and propylene \cite{2015Garcia-Lekue,2017Parreiras}. In our previous study \cite{2017Kovalenko}, we considered TPG synthesis of graphene on Ni(111) and assigned the bright features in STM images with nickel defects, i.e. with nickel atoms occupying mono- or bivacancies. However, we were not able to establish the exact structure of the objects by comparing the theoretical and experimental STM images.

In the present work, graphene was synthesized from propylene on the Ni(111) surface by both CVD and TPG methods. In both cases, identical objects similar to the defects described in \cite{2012Jacobson,2014Bianchini, 2013Patera, 2015Smerieri, 2019Carnevali, 2015Garcia-Lekue,2017Parreiras,2017Kovalenko} were found in STM images. We performed a DFT simulation of the transition from one structure (nickel in monovacancies) to another structure (nickel in bivacancies) and found that the main nickel defect in graphene is the nickel atom in the carbon bivacancy. Since nickel defects are present in graphene synthesized by both CVD and TPG, we believe that they are always formed during the graphene synthesis on the Ni(111) surface. It should also be noted that metal atoms occupying carbon bivacancies in graphene were already considered theoretically in \cite{2013WangLu, 2015Qiu, 2016Xu, 2018Zhang, 2019Mao}. Nickel defects have a positive charge and are promising from the point of view of single-atom catalysis \cite{2019Mao}.

\section{\label{sec:method}Methods}

\subsection{\label{ssec:exp}Experimental Techniques}
TPG of graphene was carried out in two stages: (i) propylene was adsorbed from 2 mm capillary  (dose of 500 Langmuir) on a purified Ni(111) surface at room temperature; (ii) sample was annealed at $500^\circ$C in UHV without propylene flux until a monolayer graphene coverage was formed. CVD of graphene was performed by supplying propylene at a pressure of $3\times 10^{-5}$\,Torr and temperature of $500^\circ$C on the clean Ni(111) surface. The clean Ni(111) surface was prepared by standard cycles, including etching by 1 keV Ar+ ions and subsequent annealing at $600^\circ$C for 15 min. All technological processes and surface analysis were performed in an ultrahigh-vacuum setup with a base pressure better than $1\times 10^{-10}$\,Torr. STM measurements were carried out at room temperature using scanning tunneling microscope GPI\,300\cite{SigmaScan}.

\subsection{\label{ssec:com}Theoretical methods}

All DFT calculations were performed in the spin-polarized version, implemented in the VASP software package \cite{1993Kresse, 1996Kresse}. Generalized gradient approximation (GGA) and exchange-correlation PBE functional \cite{1996Perdew} were applied (Van der Waals interactions were taken into account using the Grimme method \cite{2006Grimme}). The Ni(111) surface was modeled as a set of repeated 6$\times$6 cells, each containing four layers of nickel atoms, separated by a vacuum gap of 17\,{\AA}. The atomic arrangements in the lower two layers were fixed during calculations, whereas the atoms from other layers could relax. The graphene structure was optimized on the upper nickel layer so that carbon atoms were in the on-top positions (above nickel atoms) and in the fcc positions (on-top/fcc configuration). The activation barrier was calculated using the nudged elastic band (NEB) method \cite{1998NEB}, implemented in the VASP package. The reciprocal-space partition into a 6$\times$6$\times$1 grid was carried out using the G-centered k-point grid. NEB calculations were carried out applying a 3$\times$3$\times$1 k-point grid. STM images were simulated using the Hive STM program \cite{2008Vanpoucke} in the Tersoff–Hamann approximation \cite{1985Tersoff}.

\section{Results and Discussion}

\begin{figure}[h]
\begin{center}
    \includegraphics[width=\linewidth]{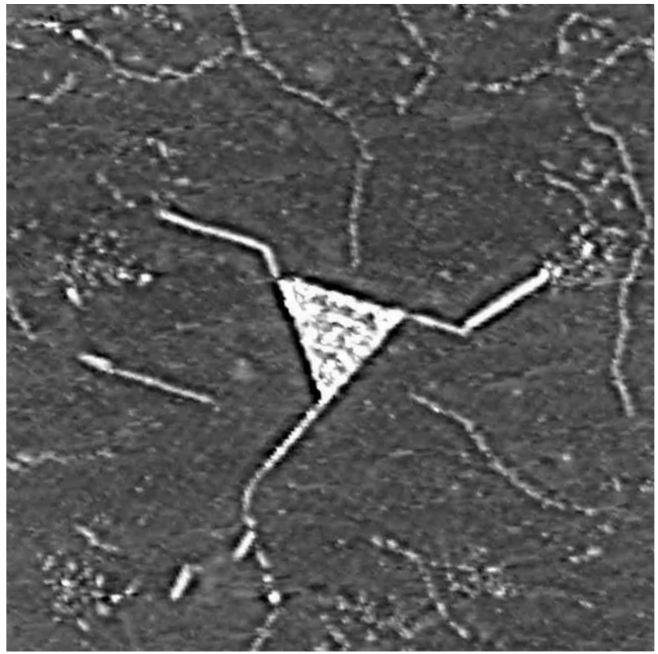}
    \caption{STM image ($936\times936$\,{\AA}$^2$, $I_{t}=0.2$ nA, $U_{s}=-1128$ mV) of the Ni(111) surface, almost completely covered with graphene. There is a two-dimensional nickel island at the center, with emerging rays in the form of one-dimensional nickel islands, transformed eventually into individual nickel atoms.}
    \label{fig1}
\end{center}
\end{figure}

Figure~\ref{fig1} shows an STM image of the Ni(111) surface almost completely covered with graphene. There is a triangular nickel island at the center of the STM image; it is surrounded with graphene located on the preceding nickel terrace. The rays emerging from the graphene island are one-dimensional residues of nickel islands, which pass eventually into separate points (nickel atoms). This Ni(111) surface structure is characteristic for TPG graphene, when heating is terminated before a continuous graphene layer is formed.

\begin{figure}[h]
\begin{center}
    \includegraphics[width=\linewidth]{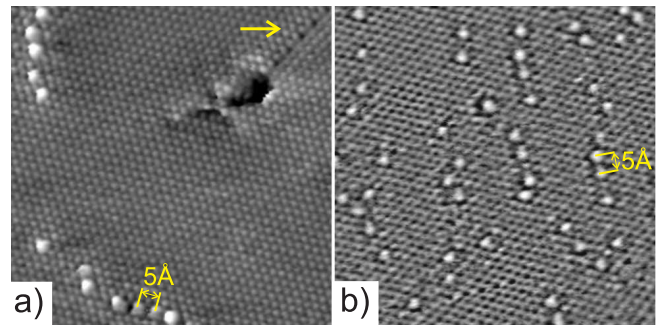}
    \caption{(a) STM image ($88\times88$\,{\AA}$^2$, $I_{t}=1.3$\,nA, $U_{s}=-7$\, mV) of the Gr/Ni(111) surface obtained by TPG. There are nickel defects on the left and a structural extended defect (indicated by arrow) on the right. (b) STM image ($88\times88$\,{\AA}, $I_{t}=2.1$\,nA, $U_{s}=-2$\,mV) of Gr/Ni(111) surface obtained by CVD. Note that the nickel defects formed by (a) TPG and (b) CVD are similar.}
    \label{fig2}
\end{center}
\end{figure}

To compare nickel and structural defects, Fig.~\ref{fig2}a presents an STM image of TPG graphene, which contains defects of both types. Nickel defects are brighter than the structural ones (see Fig. 2a). Note that these structural defects  (Fig.~\ref{fig2}a) are very rare, because graphene grows epitaxially during TPG, and structural defects are scarce in it. Similar nickel defects are observed in CVD graphene (Fig.~\ref{fig2}b). The minimum distance between the defects is about 5\,{\AA} (which makes two periods of Ni(111) lattice) for both graphene synthesis techniques.

It was shown previously \cite{2017Kovalenko} that the bright points in STM images are nickel defects: nickel atoms occupying a monovacancy (Ni-MV) or a bivacancy (Ni-BV). However, we could not reveal the exact defect structure by comparing the STM images, because Ni atoms in mono- and bivacancies yield almost identical theoretical STM images. Now, we have investigated the defect transformation from Ni-MV to Ni-BV using DFT and established the defect type.

\begin{figure}[h]
\begin{center}
    \includegraphics[width=\linewidth]{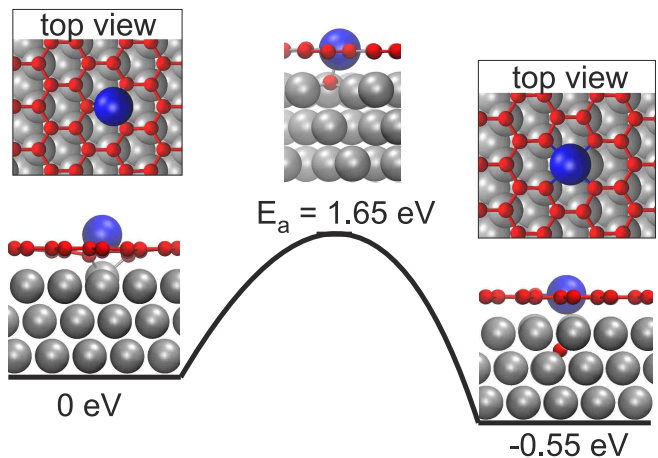}
    \caption{Energy diagram of the transition of a Ni adatom from a monovacancy to a bivacancy with carbon atom detachment from graphene and its dissolution under the first Ni(111) layer. Substrate Ni atoms, Ni atom in bivacancy, and C atoms are colored gray, blue, and red, respectively.}
    \label{fig3}
\end{center}
\end{figure}

Figure~\ref{fig3} shows the minimum-energy path for a Ni atom from the monovacancy to the bivacancy. Among two possible models of a Ni-MV defect (carbon atom vacancy in the on-top or fcc position), we chose the model, in which the Ni atom replaces the C atom in the on-top position. This model is energetically more favorable (by 0.69 eV) than that where the Ni atom occupies a vacancy in the fcc position, because the Ni adatom in the on-top position forms a bond with the substrate Ni atom located below. The defect expansion in graphene from mono- to bivacancy occurs due to the detachment of the carbon atom neighboring the vacancy and its dissolution under the first substrate layer (in the octa position). The occurrence of reaction barrier is due to the energy cost for breaking the bond between specified carbon atom and two neighboring C atoms in graphene. This energy cost is partially compensated for by the following processes: (i) formation of a bond between the distant carbon atom and six substrate atoms, (ii) formation of an additional bond between the Ni adatom and carbon atom in graphene (the Ni adatom in the Ni-BV model is bound with four C atoms rather than with three, as in the Ni-MV model), and (iii) formation of a bond between the Ni adatom and three substrate atoms. As a result of the formation of new bonds, the Ni adatom is stabilized in a bivacancy. The energy of the final state (Ni-BV) is lower, even despite the fact that it is more favorable for carbon atom to be in graphene than to be dissolved under the first Ni(111) layer. The activation energy  ($E_a$) of the Ni-MV $\rightarrow$ Ni-BV transition is 1.65 eV (see Fig.~\ref{fig3}).

The average time ($\tau$) of the Ni-MV $\rightarrow$ Ni-BV transition at a temperature $T$ can be estimated using the following formula from the transition-state theory \cite{2012Cui}:
\begin{equation}
\tau = \frac{h}{k_B T} \exp\Biggl\{\frac{E_a}{k_B T}\Biggr\}, \label{eq:1}
\end{equation}
where $h$ is the Planck's constant, $k_B$ is the Boltzmann constant, and $E_a$ is the transition activation energy. When graphene synthesis is performed at a temperature of $500^\circ$C, the average time of the Ni-MV $\rightarrow$ Ni-BV  transition is about 0.01 s. Therefore, if nickel atoms incorporate into graphene with occupation of monovacancies during the synthesis, the sample-heating time is sufficient for them to pass to a more stable state (in bivacancies).

\begin{figure}[h]
\begin{center}
    \includegraphics[width=\linewidth]{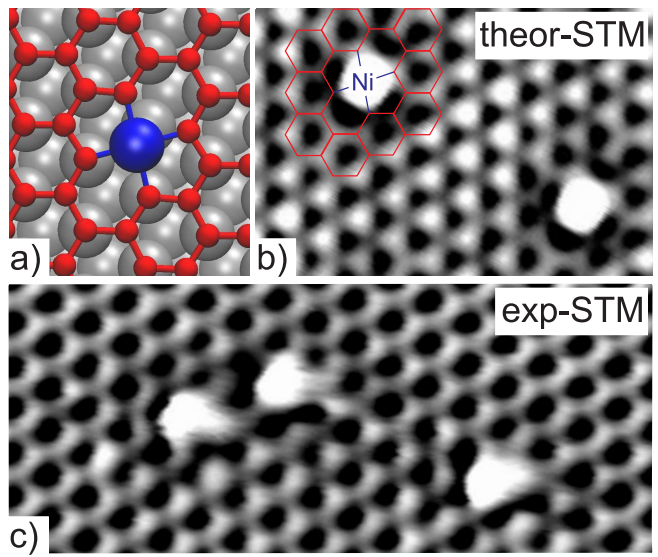}
    \caption{(a) Structure of a defect with Ni atom in graphene bivacancy on Ni(111) (substrate Ni atoms, Ni atom in bivacancy, and C atoms are colored gray, blue, and red, respectively). (b) Theoretical STM image of the defect ($U_s = -50$mV). (c) Experimental STM image ($36\times15$\,{\AA}$^2$, $I_{t}=1.4$\,nA, $U_{s}=-64$\,mV) of graphene with incorporated nickel atoms.}
    \label{fig4}
\end{center}
\end{figure}

Thus, the main type of nickel defects in Gr/Ni(111) is Ni-BV. Figure~\ref{fig4} shows the model of a Ni-BV defect (Fig.~\ref{fig4}a) and theoretical STM image (Fig.~\ref{fig4}b), which is in agreement with the experimental image (Fig.~\ref{fig4}c). The distance between the nearest Ni-BV defects cannot be smaller than two lattice periods of the Ni(111) face; otherwise, the defect structure is violated. This statement is in agreement with the experimental STM images, in which the minimum distance between defects equals to two nickel lattice periods (see Fig.~\ref{fig2}).

According to the Bader charge analysis, a Ni adatom in a graphene bivacancy on the Ni(111) surface is charged positively with lack of 0.7 electron in comparison with neutral Ni atom. Some part of the Ni adatom charge is transferred to the nearest carbon atoms, thus forming a negatively charged region around the Ni adatom. The atoms of the upper Ni(111) layer are also charged positively, and their average charge is 0.1e. The surface Ni atoms transfer their charge to carbon atoms in graphene; therefore, the charge distribution of the Ni-BV defect in graphene formed on a substrate differs from that in freestanding graphene \cite{2019Mao}.

\begin{figure}[h]
\begin{center}
    \includegraphics[width=0.85\linewidth]{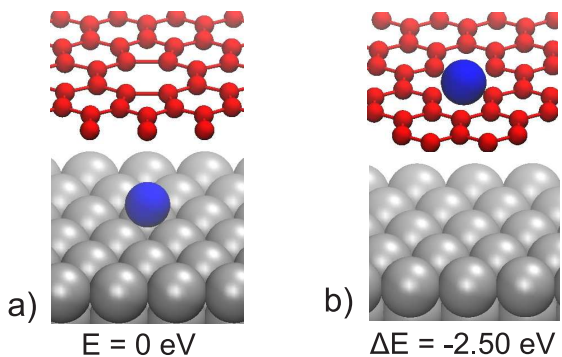}
    \caption{Structural models of graphene at a distance of 10.6\,{\AA} above the Ni(111) surface: (a) Ni adatom is on the Ni(111) surface in the fcc position and (b) Ni adatom occupies a bivacancy in graphene.}
    \label{fig6}
\end{center}
\end{figure}

We also considered models of two structures that can be formed when detaching graphene from the Ni(111) surface: (i) a Ni adatom remains on Ni(111) surface (Fig.~\ref{fig6}a) or (ii) a Ni-BV defect remains in freestanding graphene (Fig.~\ref{fig6}b). A comparison of the total energies of these two models shows that, if graphene is detached from the substrate, Ni atoms remain in graphene; this model is energetically more favorable (by 2.50 eV). Thus, the Ni adatom is bound with graphene stronger than with the Ni(111) surface atoms. Therefore, the Ni defects formed during graphene synthesis on Ni(111) can be retained when graphene is detached from the substrate, as was confirmed experimentally \cite{2015Qiu}.

\section{Conclusions}

The nickel defects formed in graphene synthesized on Ni(111) by CVD and TPG methods are identical. When analyzing the Ni(111) surface during TPG, we were able to obtain an STM image of intermediate structures and establish that nickel atoms remain in graphene (looking like bright points in the image). According to DFT, the Ni adatom in graphene is more stable in a carbon bivacancy rather than in a monovacancy. Our calculations show that the configuration with a Ni atom in monovacancy transforms into the configuration with a Ni atom in bivacancy due to the dissolution of one carbon atom into the nickel bulk over a period of about 0.01 s at a graphene synthesis temperature. We have found also that the Ni adatom is charged positively, and there is a negatively charged region around it. Such positively charged centers in graphene have the potential for single-atom catalysis. The Ni adatom in a bivacancy is more strongly bound with graphene rather than with the substrate; therefore, when graphene is detached, it remains in graphene bivacancy. Ni-BV centers on the Ni(111) surface may also be of interest. The presence of these centers in graphene on a substrate opens ways to study single-atom catalysts by STM and atomic force microscopy, observing in situ transformations of individual molecules.

\section{Acknowledgements}

This study was supported by the Russian Foundation for Basic Research, project no. 16-29-06426. We also thank the Joint Supercomputer Center of RAS for providing the computing power.

\bibliography{Ni_defect_Gr_arXiv}

\end{document}